\documentclass[aps,prb,twocolumn,floatfix,showpacs,superscriptaddress]{revtex4}

\usepackage{amssymb}
\usepackage{epsfig}
\usepackage{amsmath}
\usepackage{color}

\def \Grenoble{Laboratoire National des Champs Magn\'etiques Intenses, CNRS-UJF-UPS-INSA, 38042 Grenoble, France}
\def \Toulouse{Laboratoire National des Champs Magn\'etiques Intenses, CNRS-UJF-UPS-INSA, 31400 Toulouse, France}

\begin{document}

\title{High field magneto-transmission investigation of natural graphite}
\author{N. \surname{Ubrig}}\affiliation{\Toulouse}
\author{P. \surname{Plochocka}}\affiliation{\Grenoble}
\author{P. \surname{Kossacki}}\affiliation{\Grenoble}
\author{M. \surname{Orlita}}\affiliation{\Grenoble}
\author{D. K. \surname{Maude}}\affiliation{\Grenoble}
\author{O. \surname{Portugall}}\affiliation{\Toulouse}
\author{G. L. J. A. \surname{Rikken}}\affiliation{\Toulouse}\affiliation{\Grenoble}

\date{\today}

\begin{abstract}
Magneto-transmission measurements in magnetic fields in the range $B=20-60$~T have been performed to probe the $H$ and
$K$-point Landau level transitions in natural graphite. At the $H$-point, two series of transitions, whose energy
evolves as $\sqrt{B}$ are observed. A reduced Slonczewski, Weiss and McClure (SWM) model with only two parameters to
describe the intra-layer ($\gamma_0$) and inter-layer ($\gamma_1$) coupling correctly describes all observed
transitions. Polarization resolved measurements confirm that the observed apparent splitting of the $H$-point
transitions at high magnetic field cannot be attributed to an asymmetry of the Dirac cone.
\end{abstract}

\pacs{81.05.uf, 71.70.Di, 78.20.Ls, 78.30.-j}

\maketitle


Graphite consists of Bernal stacked sheets of hexagonally arranged carbon atoms.  The weak coupling between the layers
transforms the single graphene layer, which is a gapless semiconductor with a linear dispersion, into a semimetal with
electron and hole puddles along the $H-K-H$ edge of the hexagonal Brillouin zone.\cite{Wallace47} In a magnetic field
the electronic structure of graphite is accurately described by the Slonczewski, Weiss and McClure (SWM) band structure
calculations, \cite{Slonczewski58,McClure60} which require seven tight binding parameters $\gamma_0,..,\gamma_5,
\Delta$ to define the interaction energy of the carbon atoms in the graphite lattice. The SWM model has been
extensively verified using Shubnikov de Haas, de Haas van Alphen, thermopower and magneto-reflectance
experiments.~\cite{Soule58,Soule64,Woollam70,Woollam71a,Schneider09,Zhu10,Williamson65,Schroeder68} Carriers at the
$H$-point behave as relativistic Dirac Fermions with a linear dispersion as in graphene. Magneto-absorption has been
used to perform Landau level spectroscopy of carriers at the $H$-point ($k_z=0.5$) and $K$-point ($k_z=0$) in both
natural graphite and highly ordered pyrolytic graphite (HOPG).\cite{Doezema79,Orlita08,Orlita09,Chuang09}


At the $H$-point, transitions with a characteristic $\sqrt{n B}$ magnetic field dependence of their energy are
observed, which are identical to the transitions at the $K$ and $K'$-points observed in graphene. For this reason, we
refer to this series as ``graphene-like'' although we stress that here the series arises from the $H$-point transitions
of perfect bulk graphite. In addition, a second weaker series of transitions with a characteristic $\sqrt{n B}$
magnetic field dependence of their energy are observed. These transitions are absent in graphene, in fact they
correspond to dipole forbidden transitions of the ``graphene-like'' series. However, this series correspond to dipole
allowed transitions in graphite due to the complicated band structure at the $H$-point. \cite{Nakao76,Toy77,Orlita08}
We refer to these transitions as the ``graphite-like'' series, since they are absent in graphene. For the $K$-point
there is evidence of a splitting of the transitions which has been attributed to electron-hole
asymmetry.\cite{Chuang09}

Here we report magneto-optical absorption measurements to probe the evolution of the $K$ and $H$ point transitions in
magnetic fields up to $60$~T. This extends previous work\cite{Doezema79,Orlita08,Orlita09,Chuang09} to higher magnetic
fields and more importantly to higher energies. In particular, the use of near visible radiation facilitates the
implementation of polarization resolved measurements. The observed transmission spectra are dominated by the Dirac-like
series of transitions from the $H$-point. All the observed transitions can be assigned, and the magnetic field
evolution reproduced, using a reduced SWM model with two tight binding parameters $\gamma_0$ and $\gamma_1$.
Polarization resolved measurements confirm that the observed splitting of the $H$-point ``graphene-like'' transitions
is not linked to the asymmetry of the Dirac cone, which is anyway irrelevant at the $H$-point within the SWM model.
Upon closer examination, the splitting resembles rather an avoided level crossing, while nevertheless remaining
unexplained.


Thin samples for the transmission measurements were prepared by exfoliating natural graphite. The average thickness of
the graphite layers remaining on the foil was estimated to be $\simeq100$~nm.\cite{Orlita08} The measurements were
performed up to $34$~T at the \emph{dc} resistive magnet laboratory in Grenoble and up to $60$~T at the pulsed magnetic
field laboratory in Toulouse. For the absorption measurements a tungsten halogen lamp was used to provide broad
spectrum in the visible and near infra-red range. The absorption was measured in the Faraday configuration in which
$k$, the wave propagation vector is parallel to the magnetic field, $B$. The $c$-axis of the graphite sample was
parallel to magnetic field.  A nitrogen cooled InGaAs photodiode array coupled to a spectrometer collected the
transmitted light from the sample in the spectral range $850-1600$~nm, \emph{i.e.} energies of $0.8-1.5$~eV. For the
pulsed field measurements the exposure time was limited to $2$~ms in order to limit variations in the magnetic field
during acquisition. Thirty spectra were taken during a $60$~T shot so that in principle a complete magnetic field
dependence can be acquired in a single shot. The magnetic field was systematically measured using a calibrated pick-up
coil. Since the absorption lines in this energy range are weak all the spectra were normalized by the zero field
transmission to produce a differential transmission spectra.

\begin{figure}[]
\begin{center}
\includegraphics[width= 8.5cm]{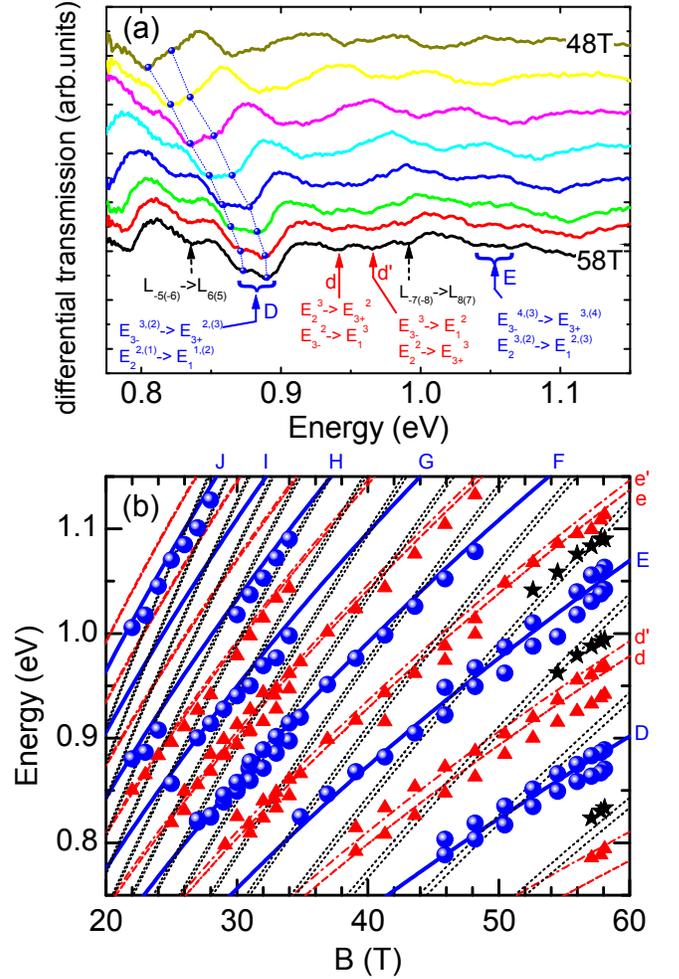}
\end{center}
\caption{(Color online) (a) Typical differential magneto-transmission spectra of natural graphite measured at magnetic
fields in the range $48-58$~T at $T=4.2$~K. (b) Magnetic field dependence of the observed transitions assigned as
follows: $H$-point, graphene series (blue balls), graphite specific series (red triangles); $K$-point (black stars).
The lines are calculated energies of the dipole allowed $H$-point (solid, dotted, and, dot-dashed lines) and $K$ point
(dashed lines) transitions as described in the text.} \label{fig1}
\end{figure}

Typical differential magneto-absorption spectra measured at $T=4.2$~K for magnetic fields $48-58$~T are shown in
Figure~\ref{fig1}(a). All spectra show a number of absorption lines which can be assigned to dipole allowed transitions
at the $H$ and $K$ points. The energetic position of the observed absorption lines is plotted as a function of the
magnetic field in Figure~\ref{fig1}(b). In order to assign the transitions we first calculate the energy of the dipole
allowed transitions ($\Delta n=\pm 1$) at the $H$ and $K$-points using a greatly simplified SWM model with only two
parameters $\gamma_0$ and $\lambda \gamma_1$ to describe the intra- and inter-layer
coupling.\cite{Partoens2006,Partoens2007,Chuang09,Koshino08,Orlita09} Here $\lambda=2 \cos(\pi k_z)$ and $k_z$ is the
momentum perpendicular to the layers. This corresponds to treating graphite as a series of graphene bi-layers whose
effective coupling depends on $k_z$. The magneto-optical response is dominated by the singularities in the joint
density of initial and final states which occur at the $K$-point ($\lambda=2$) and $H$-point ($\lambda=0$). The energy
spectrum of the Landau levels using the effective bilayer model is then given by,
\begin{multline}\label{Bilayer}
E^{n}_{3\pm}=\pm\frac{1}{\sqrt{2}}\left[(\lambda\gamma_1)^2+(2n+1)\varepsilon^2\right.\\
\left.-\sqrt{(\lambda\gamma_1)^4 +2(2n+1)\varepsilon^2(\lambda\gamma_1)^2+\varepsilon^4}\right]^{1/2},
\end{multline}
where $\varepsilon=\tilde{c}\sqrt{2e\hbar B}$ is the characteristic magnetic energy, $\tilde c=\sqrt3 e a_0
\gamma_0/2\hbar$ is the Fermi velocity, $a_0=0.246$~nm is the lattice constant in the $ab$ plane and $\pm$ labels the
electron and hole Landau levels respectively. At the $H$-point, equation~(\ref{Bilayer}) reduces to the Landau level
spectrum of graphene with $E^{n}_{3\pm}=\pm \tilde c \sqrt{2 e \hbar B n}$.

\begin{figure}[]
\begin{center}
\includegraphics[width= 8.5cm]{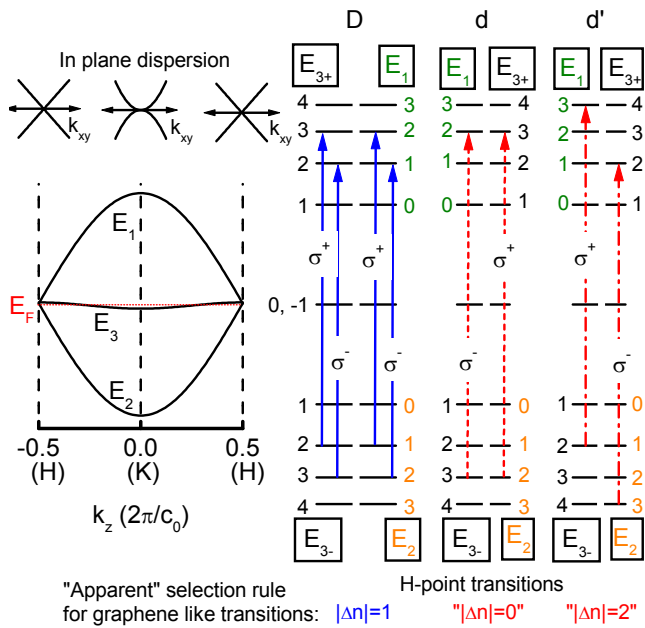}
\end{center}
\caption{(Color online) (left) Band structure of graphite along the hexagonal $H-K-H$ axis. (right) Dipole allowed
transitions ($\Delta n=1$) in a magnetic field at the $H$-point corresponding the $E_{3-}\rightarrow E_{3+}$ (and
$E_{2}\rightarrow E_{1}$) ``graphene-like'' transitions (labeled $D$), and the `mixed' $E_{3-}\rightarrow E_{2}$ and
$E_{2}\rightarrow E_{3+}$ ``graphite-like'' transitions (labeled $d$ and $d'$). The circular polarization required to
excite each transition is indicated using the convention that $\sigma^+$ polarization corresponds to $\Delta n=+1$.}
\label{fig2}
\end{figure}

The bi-layer model is expected to be almost exact at the $H$-point since the effect of trigonal warping ($\gamma_3$)
vanishes and analytic expressions for the Landau levels can be easily obtained within the SWM model by diagonalizing
the Hamiltonian.\cite{Nakao76} However, the situation is complicated by the presence of the $E_1$ and $E_2$ bands (see
Fig.\ref{fig2}), which are almost degenerate with $E_3$ at the $H$-point (energy splitting $\Delta\simeq-0.007$~eV). In
a magnetic field, neglecting the two exceptional $E_{3}$ Landau levels ($n=0,-1$), this gives rise to a second Landau
level spectrum, $E^{n}_{1,2}=E^{n+1}_{3\pm}$ where $n=1,2,3,..$ which is exactly degenerate with the $E_{3\pm}$ ladder.
\cite{Nakao76,Orlita08,Orlita09} The Landau level spectrum at the $H$-point is shown schematically in Fig.~\ref{fig2}
where we indicate all possible dipole allowed $E^{2(3)}\rightarrow E^{3(2)}$ transitions as an example. The graphene
like transition $E^{3(2)}_{3-}\rightarrow E^{2(3)}_{3+}$ (labeled $D$) have the same energy as the
$E^{2(1)}_{2}\rightarrow E^{1(2)}_{1}$ transitions which have a quantum number $n$ which is lower by one. The circular
polarization of the light required to excite each transition is indicated and we have adopted the convention that
$\sigma^+$ polarization corresponds to $\Delta n=+1$. The transition labeled $d$ and $d'$ are specific to graphite
(``graphite-like'' series). Transition $d$ is the dipole allowed ($|\Delta n|=1$) degenerate `mixed' transitions
$E^{3}_{3-} \rightarrow E^{2}_{1}$ and $E^{2}_{2} \rightarrow E^{3}_{3+}$ which correspond to (are exactly degenerate
with) dipole forbidden ($\Delta n=0$) transitions of the graphene series. Transition $d'$ shows dipole allowed
($|\Delta n|=1$) degenerate `mixed' transitions $E^{2}_{3-} \rightarrow E^{3}_{1}$ and $E^{3}_{2} \rightarrow
E^{2}_{3+}$ which correspond to (are exactly degenerate with) dipole forbidden transitions $|\Delta n|=2$ of the
graphene series. Note, that while we cannot exclude the presence in our sample of decoupled graphene layers, with
transitions degenerate with the ``graphene-like'' series, the overwhelming contribution of graphite to the transmission
is demonstrated by the observed strength of the ``graphite-like'' series.

The energy of the dipole allowed optical transitions, calculated using Equation(\ref{Bilayer}) with the tight binding
parameters, $\gamma_0=3.15$~eV ($\tilde c=1.02 \times 10^6$ ~m.s$^{-1}$) and $\gamma_1=0.375$~eV determined from
magneto-absorption measurements at lower magnetic fields, \cite{Orlita09} are plotted as a function of the magnetic
field in Fig.~\ref{fig1}(b) (solid and broken lines).  The $H$-point transitions depend only on the parameter
$\gamma_0$ and evolve always as $\sqrt{B}$. The $K$-point transition depend also on the inter-layer coupling $\lambda
\gamma_1$, and therefore, as can be seen from Equation(\ref{Bilayer}), evolve linearly at low energy
($\varepsilon\ll\lambda\gamma_1$) before increasing as $\sqrt{B}$ at high energies ($\varepsilon\gg\lambda\gamma_1$).
At the $K$-point $\lambda \gamma_1=0.75$~eV so that we are in the intermediate regime where dependence is somewhere
between linear and $\sqrt{B}$.

The agreement between the reduced two-parameter SWM model and experiment in Figure~\ref{fig1}(b) is remarkable,
especially taking into account that there are no adjustable parameters. The $H$-point transitions $E^{2(3)}\rightarrow
E^{3(2)}$, are labeled as in Fig.\ref{fig2}. Mainly $H$-point transitions are observed, notably the ``graphene-like''
series $E_{3-} \rightarrow E_{3+}$ and $E_{2} \rightarrow E_{1}$ (thick blue solid lines labeled with upper case
letters) together with the weaker $E_{3-} \rightarrow E_{1}$ and $E_{2} \rightarrow E_{3+}$ transitions (red dashed and
dot-dashed lines labeled with lower case letters). The $K$-point transitions, shown as black dotted lines are only
observed directly at high magnetic fields. For completeness, for the $K$-point transitions, we have included
phenomenologically the electron-hole asymmetry as suggested in Refs.[\onlinecite{Henriksen08, Chuang09}] by using a
different Fermi velocity $\tilde c_e=1.098 \times 10^6$~ms$^{-1}$ and $\tilde c_h=0.942 \times 10^6$~ms$^{-1}$ for the
electrons and holes respectively. These values are slightly different from those used in Ref.\onlinecite{Chuang09} in
order to have the same ``average'' value of $\gamma_0=3.15$~eV. While the electron-hole asymmetry was clearly seen in
measurements at low magnetic field,\cite{Chuang09} the phenomenological asymmetry splitting introduced in
Ref.[\onlinecite{Chuang09}] decreases rapidly with increasing quantum number, and is probably too small to be seen in
our high magnetic field data (the lowest energy $K$-point transition seen is $n=5$, labeled $L_{-5 (-6)}\rightarrow
L_{6 (5)}$ in Figure~\ref{fig1}(a)).

The ``graphene-like'' series unexpectedly shows what looks at first sight to be a splitting, which is puzzling since
such a splitting is completely absent in magneto-transmission measurements on graphene.\cite{Plochocka08} This apparent
splitting is clearly seen in the $E^{n (n+1)}_{3-} \rightarrow E^{n+1 (n)}_{3+}$ transitions ($n=2,3,4$) labeled
$D$,$E$ and $F$ in Fig.\ref{fig1}. However, a closer inspection of the magnetic field evolution of the energy of the
strong $E^{2 (3)}_{3-} \rightarrow E^{3 (2)}_{3+}$ in Fig.~\ref{fig1}(b) (transition $D$) indicates that the calculated
transition fits better to the low energy feature at low fields ($B<50$~T) before fitting better to the high energy
feature at high fields ($B>54$~T). This is suggestive of an avoided level crossing rather than a splitting. This
hypothesis is supported by the absorption spectra in Fig.~\ref{fig1}(a), where it is clearly seen that the $E^{2
(3)}_{3-} \rightarrow E^{3 (2)}_{3+}$ doublet (transition $D$) consists of a stronger low energy transition at low
magnetic fields which switches to a stronger high energy feature at high fields, \emph{i.e.} the two lines anti-cross.
The origin of this behavior remains to be elucidated. However, this cannot be due to inhomogeneity of the sample. A
slightly different Fermi velocity for different regions would simply lead to an increased splitting with increasing
magnetic field.

\begin{figure}[]
\begin{center}
\includegraphics[width= 5.5cm]{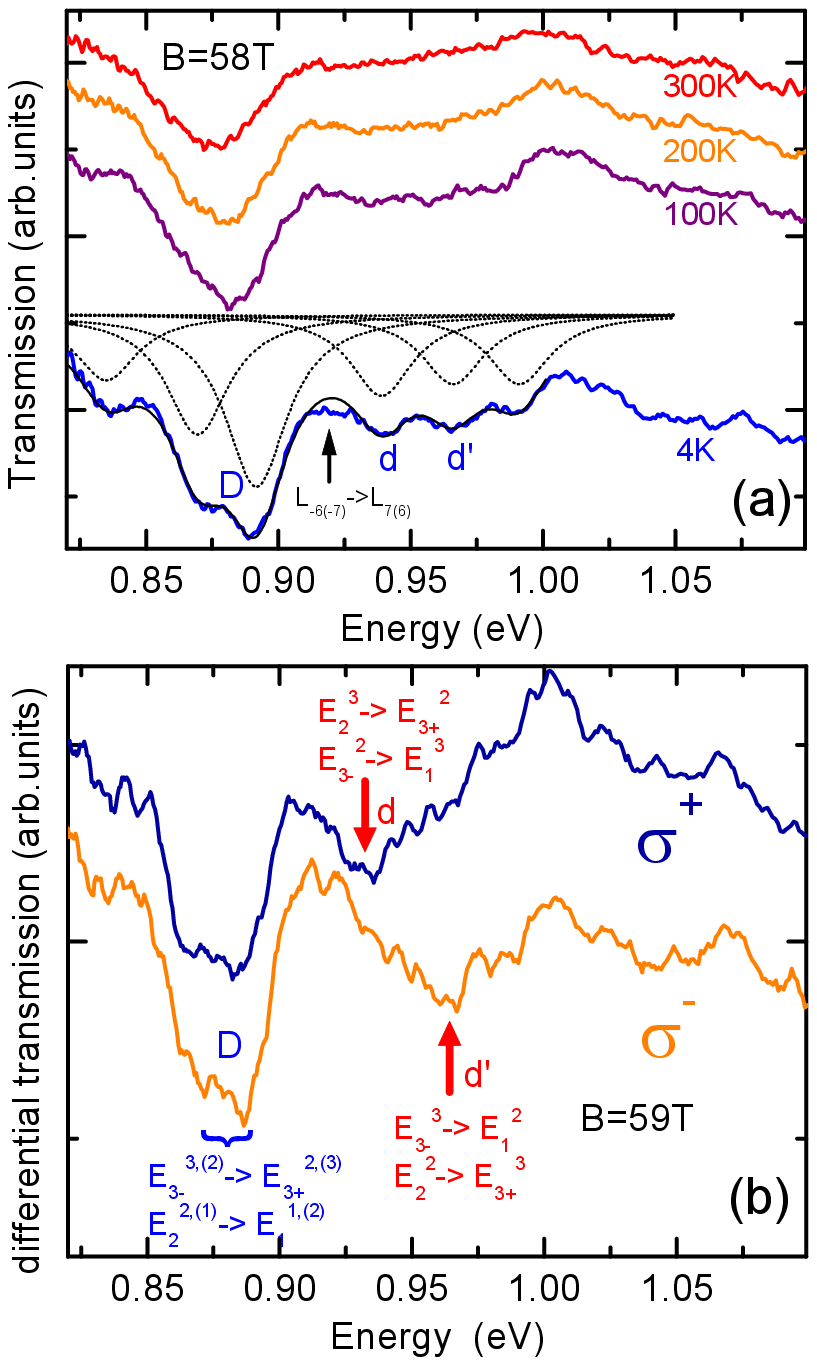}
\end{center}
\caption{(Color online) (a) Differential magneto-transmission spectra of natural graphite measured at $B=58$~T for
different temperatures. The thin black lines are a fit to the $T=4$~K spectra assuming a Lorentzian line shape.  (b)
Polarization resolved spectra measured at $B=59$~T and $T= 4.2$~K for the transitions $D$, $d$ and $d'$ sketched
schematically in Fig.~\ref{fig2}. The polarization ($\sigma^+$ or $\sigma^-$) has been arbitrarily assigned to a given
magnetic field direction.} \label{fig3}
\end{figure}

Figure~\ref{fig3}(a) shows differential absorption spectra measured at $B=58$~T for different temperatures in the range
$4-300$~K. A temperature of $100$~K is already sufficient to suppress the apparent splitting of the $E^{2 (3)}_{3-}
\rightarrow E^{3 (2)}_{3+}$ transition, although the transition itself, while weakening slightly, remains clearly
visible even at room temperature. The $T=4$~K spectra have been fitted using a Lorentzian line shape of full width at
half maximum (FWHM) of $28.5$~meV for all transitions. The result of the fit (solid thin black line in
Fig.\ref{fig3}(a)) describes the data extremely well. The individual Lorentzians, for each transition are shown as
dotted lines. Clearly, the broadening of the transitions is comparable to the energy separation so that the absorption
is only weakly modulated. Note that the disagreement between the data and the fit around $0.92$~eV is probably a
signature of the ``missing'' L$_{-6(-7)}\rightarrow$L$_{7(6)}$ $K$-point transition in Fig.\ref{fig1} which could not
be assigned from the raw data. Keeping all other parameters constant, increasing the broadening of the Lorentzians
produces a reasonable fit to the higher temperature data, with the exception of the $E^{2 (3)}_{3-} \rightarrow E^{3
(2)}_{3+}$ transition. A reasonable fit to this transition at higher temperatures requires, in addition to a thermal
broadening, that the amplitude of the two Lorentzian components be changed for which we see no physical justification.
We therefore conclude that thermal broadening alone cannot explain the observed temperature dependence of the $E^{2
(3)}_{3-} \rightarrow E^{3 (2)}_{3+}$ transition .

>From a theoretical point of view a splitting of the $E^{2 (3)}_{3-} \rightarrow E^{3 (2)}_{3+}$ transition (or the
degenerate $E^{1 (2)}_{2} \rightarrow E^{2 (1)}_{1}$) transition) is not expected since even in the full SWM model. The
effect of $\gamma_3$ (trigonal warping) vanishes at the $H$-point so that the energy levels are determined only by
$\gamma_0$. The non vertical inter layer coupling term $\gamma_4$, which induces electron-hole asymmetry, plays no
role. This can be verified experimentally using the polarization resolved absorption measured at $59$~T shown in
Fig.~\ref{fig3}(b) which focuses on transitions $D$, $d$ and $d'$. There is no difference between $\sigma^+$ and
$\sigma^-$ spectra for $E^{2 (3)}_{3-} \rightarrow E^{3 (2)}_{3+}$ (transition $D$) confirming that the apparent
doublet cannot under any circumstances be assigned to electron-hole asymmetry. In contrast, the `mixed' $E_{3-}
\rightarrow E_{1}$ and $E_{2} \rightarrow E_{3+}$ transitions ($d$ and $d'$) show a marked dependence on the circular
polarization with one of the transitions almost vanishing with either $\sigma^+$ or $\sigma^-$ excitation. Using the
polarization selection rules sketched in Fig.\ref{fig2} this can be explained provided one of the inter-band
transitions ($E_{3-} \rightarrow E_{1}$ or $E_{2} \rightarrow E_{3+}$) dominates. However, as in our experiment the
sense of the circular polarization has been arbitrarily assigned to a given magnetic field direction it is
unfortunately not possible to know which transition prevails.

While trigonal warping plays no role at the $H$-point because $\gamma_3$ always enters the SWM Hamiltonian as $\gamma_3
\cos(\pi k_z)$, close to the $H$-point it can lead to magnetic breakdown producing a splitting of levels in the Landau
level structure which could possibly be observed in magneto-optical spectra at the $H$-point.\cite{Nakao76} This
originates from an anti-crossing of Landau levels from the $E_3$ band with Landau levels from the $E_1$ or $E_2$ bands.
The repulsion occurs due to the interaction caused by $\gamma_3$ provided the Landau levels originate from the same
submatrix (of the three possible) of the magnetic Hamiltonian. In contrast to the $K$-point, where trigonal warping
induced magnetic breakdown occurs only at low magnetic fields, close to the $H$-point magnetic breakdown takes place
for all magnetic field strengths. An additional complication at very high magnetic fields ($B\approx70$~T) is the
predicted magnetic field induced transition of semi-metallic graphite to a zero gap semiconductor due to the crossing
of the $n=0$ Landau level at the $K$-point and the $n=-1$ Landau level at the $H$-point.\cite{Nakao76} Further
measurements at higher magnetic fields are planned to clarify these issues.


In conclusion, magneto-transmission measurements have been used to probe the $H$ and $K$-point Landau level transitions
in natural graphite. In the magnetic field range investigated, the spectra are dominated by transitions at the
$H$-point. A ``graphene-like'' series together with a series of transitions exclusive to graphite are observed. We
stress that both series arise from dipole allowed transitions at the $H$-point of \emph{perfect bulk graphite}, and do
not require the presence of decoupled graphene layers or decoupled bilayers in the sample. A reduced SWM model with
only two parameter $\gamma_0$ and $\gamma_1$ correctly describes all observed transitions. Polarization resolved
measurements (i) confirm that the apparent splitting of the ``graphene-like'' series at high magnetic field cannot be
attributed to an asymmetry of the Dirac cone and (ii) suggest that the matrix elements connecting $E_{3+} \rightarrow
E_1$ and $E_{3-} \rightarrow E_2$ are very different.

\acknowledgements{This work has been partially supported by ANR contract PNANO-019-06, Euromagnet II and grant GACR
P204/10/1020. The authors thank Sylvie George and the LNCMI machine shops for technical support. Two of us (P.P. and
P.K.) are financially supported by the EU under FP7, contract no. 221249 `SESAM' and contract no. 221515 `MOCNA'
respectively.}


\end{document}